\documentclass{fizik}
\usepackage{hyperref}
\hypersetup{
colorlinks=true,
urlcolor=blue,
citecolor=blue}
\usepackage[all]{xy,xypic}
\usepackage{amsfonts,amssymb,amsmath,amsgen,amsopn,amsbsy,theorem,graphicx,epsfig}
\usepackage{eufrak,amscd,bezier,latexsym,mathrsfs,enumerate,multirow}
\usepackage[utf8]{inputenc}\usepackage[english]{babel}
\usepackage[dvipsnames]{xcolor}
\usepackage[pagewise]{lineno}

\usepackage{subfig}
\usepackage{float}

\yil{}
\vol{}
\fpage{}
\lpage{}
\doi{}

\title{Scintillation Timing Characteristics of Common Plastics for Radiation Detection Excited With 120 GeV Protons\\}

\author[BILKI et al.]{
\textbf{Burak BILKI$^{1,2}$, 
Nilay BOSTAN$^{2}$,
Ohannes Kamer K{\"O}SEYAN$^{2}$,
Emrah TIRA{\c S}$^{2,3}$,
James WETZEL$^{2,4}$\thanks{james-wetzel@uiowa.edu}}\\
$^{1}$Department of Mathematics, Beykent University, Istanbul, Turkey \\
$^{2}$Department of Physics and Astronomy, The University of Iowa, Iowa City, Iowa, USA \\
$^{3}$Department of Physics and Astronomy, Iowa State University, Ames, Iowa, USA \\
$^{4}$Department of Physics, Augustana College, Rock Island, Illinois, USA \\
\\ [1.8em]

}

\newcommand{\bc}{\begin{center}}
\newcommand{\ec}{\end{center}}

\numberwithin{equation}{section}

\renewcommand{\phi}{\varphi}

\begin{document}

\maketitle

\begin{abstract}	The timing characteristics of scintillators must be understood in order to determine which applications they are appropriate for.  Polyethylene naphthalate (PEN) and polyethylene teraphthalate (PET) are common plastics with uncommon scintillation properties.  Here, we report the timing characteristics of PEN and PET, determined by exciting them with 120 GeV protons.  The test beam was provided by Fermi National Accelerator Laboratory, and the scintillators were tested at the Fermilab Test Beam Facility.  PEN and PET are found to have dominant decay constants of 34.91 ns and 6.78 ns, respectively. 

\end{abstract}

\section{Introduction}

		After observing that common plastics polyethylene naphthalate (PEN) and polyethylene teraphthalate (PET) have very good scintillation properties \cite{nakamura2011evidence,nakamura2013blended}, and reasonable tolerance to gamma radiation exposure \cite{JWandET}, interest in using PEN and PET in scintillation applications has grown.  Of note is their potential application in collider physics experiments, where huge amounts of active media are required in very large detectors.  These plastics' low cost and ease of manufacture are attractive properties.  However, for these applications, scintillators need to be relatively fast and bright. 
		
		In order to study their properties for future applications, similar samples of PEN and PET were prepared for testing at Fermilab Test Beam Facility (FTBF) by exciting them with  120 GeV protons.  The proton beam at FTBF produces $5\times 10^5$ protons per spill with a spot size of 6 mm. The experiment was designed to measure the decay constants of PEN and PET.  The decay constant is the time it takes for the scintillator to diminish to 1/e of its maximum signal amplitude.  As a particle traverses a scintillator, the particle excites the molecules within the material, which then radiate photons during de-excitation. 
		
		Timing is important for experiments like those at CERN, where the LHC \cite{evans2008lhc} has a design collision frequency of 40 MHz, or one collision every 25 ns.  Considering the potential for unmanageable pileup, or the excessive number of proton-proton collisions per beam bunch crossing, a scintillator must dispatch its entire signal ideally within, in the case of LHC experiments, 25 ns.
		
		In addition, PEN was recently identified as a relatively efficient wavelength-shifter for vacuum ultraviolet (VUV) photons and found an immediate implementation at the CERN ProtoDUNE-DP \cite{DP} experiment. 30 out of 36 photomultiplier tubes (PMTs) of ProtoDUNE-DP have PEN sheets installed on top of their windows, and the PMTs have been operating in liquid argon successfully since Summer 2019 \cite{dualPhase}. Studying the scintillation properties of such common plastics under various test conditions will have critical importance for a wide range of implementations in the future. 

\section{Experimental Setup}
		Three samples each of PEN and PET were prepared for testing.  The PEN sample was cut to a square 100 mm x 100 mm x 1 mm from an 8.5" x 11" sheet of Teijin Plastics Scinterex using an end mill.  The edges were polished with a polishing wheel.  The PET sample was delivered in 100 mm x 100 mm x 1 mm squares from Goodfellow.com, and the edges were polished with a polishing wheel. 
		
\begin{figure}[h!]
			\centering
				\includegraphics[width=0.8\textwidth]{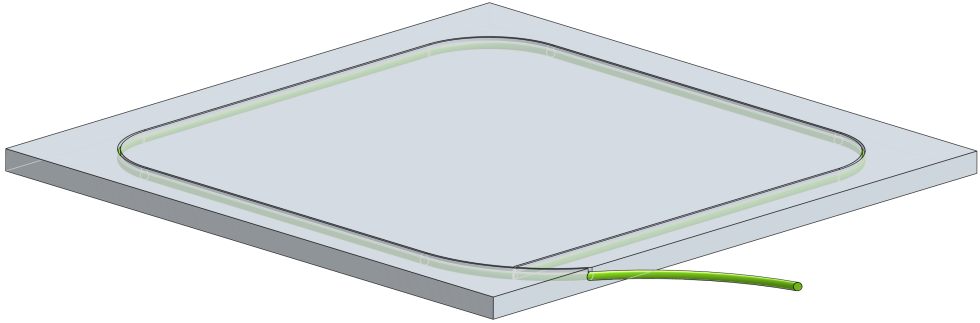}
				\caption{Scintillator tile with keyhole groove and Y11 fiber.}
				\label{fig:Groove}
		\end{figure}

\begin{figure}[]
			\begin{centering}
				\subfloat[]{\includegraphics[width=0.40\textwidth]{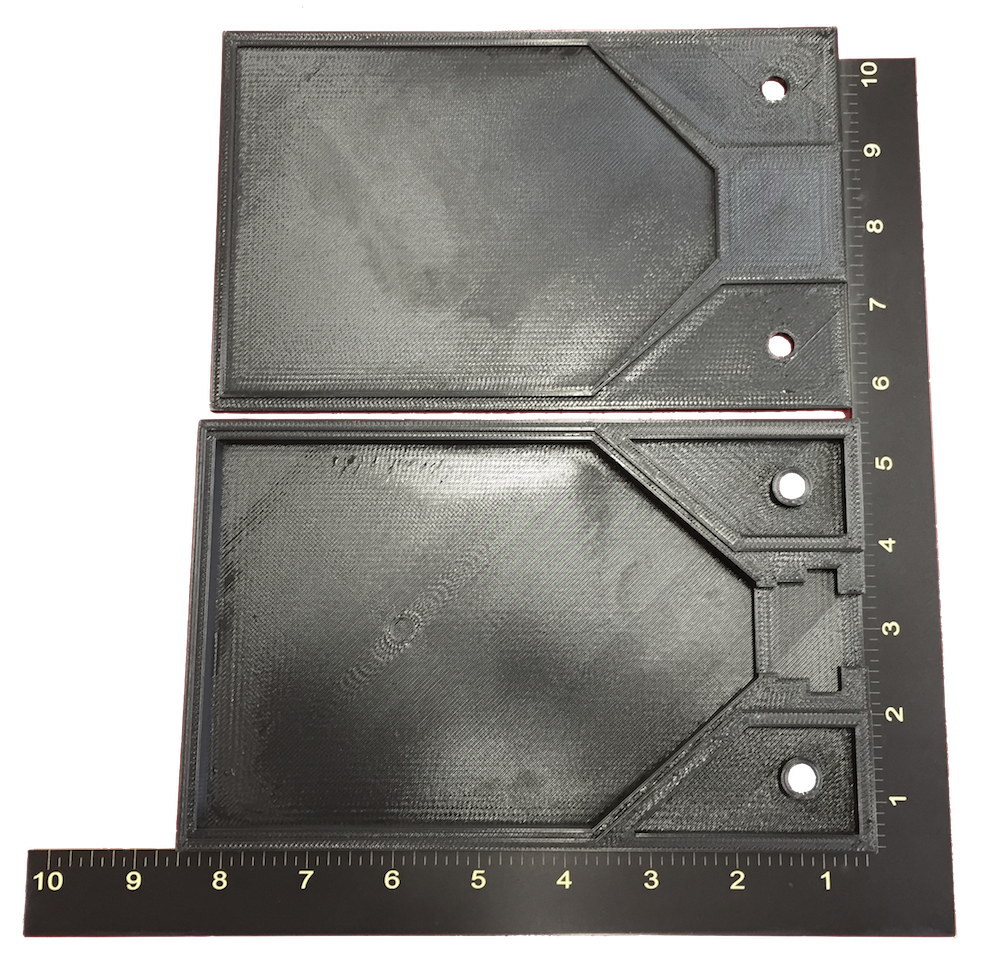}}
				\qquad
				\subfloat[]{\includegraphics[width=0.5\textwidth]{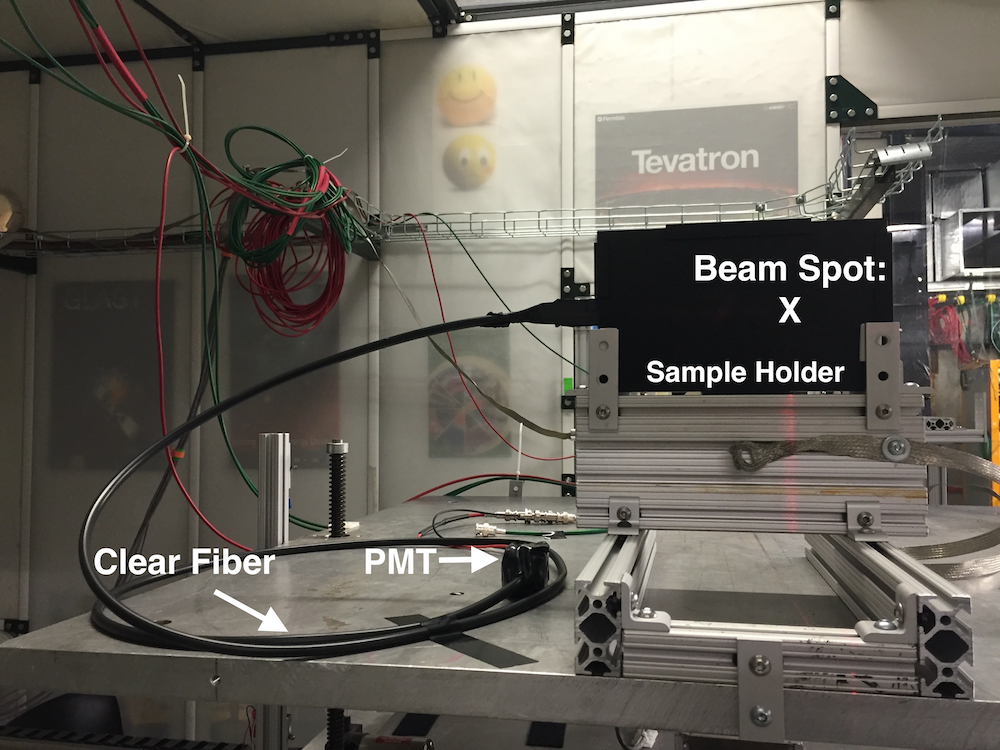}}
				\caption{3D printed frame for holding the scintillators to be tested (a) and experimental setup in the beam area showing the sample holder, the clear fiber, and the PMT (b).}
				\label{figure:PenAnalysisSetup}
			\end{centering}
		\end{figure}

		Each tile had a `keyhole' groove machined into its surface around the perimeter to hold an optical fiber.  A `Y11' Kuraray wavelength shifting (WLS) fiber was inserted into the groove, Fig \ref{fig:Groove}.  An injection molded connector was glued to the end of the fiber, and the fiber edge was polished.  Each tile was placed into a 3D printed housing, Fig \ref{figure:PenAnalysisSetup}, and the WLS fiber was coupled to a clear fiber.  The clear fiber was connected to a clear acrylic `cookie' for coupling to an R7600 PMT for readout.

		The signals recorded from the R7600 PMT were fed into a Tektronix TDS 5034 digital oscilloscope, and events were recorded in a Tektronix proprietary `.wfm' format , using the `Fast Frame' mode of the oscilloscope.  This allowed the recording of twenty-five thousand events to be saved in a single file, which we could later analyze.  The events were triggered in the central 1 cm x 1 cm area of the tiles by the coincidence of two external scintillator paddle signals.
		
	\section{Experimental Results}	
		The twenty-five thousand events were extracted from the .wfm format and ntupled using the ROOT framework \cite{antcheva2011root}, collating all events into a single ROOT file.  The pedestal was subtracted from each event, and the amplitudes of the event waveforms were averaged using ROOT's TProfile function in order to create an average waveform profile.  The TProfile function averages the contents of each bin over the twenty-five thousand events, thereby providing a very nice picture of the scintillator's photoluminescence and decay profile.  
		
		Generally, scintillators have a \emph{fast} and \emph{slow} component to their time-dependent luminescence profile, corresponding respectively to its recombination centers and electron traps.
		
		The waveform profile of a scintillator can therefore be modeled as a combination of slow and fast components \cite{swiderski2014measuring}:
		
		\begin{equation}\label{2exp}
			\frac{A}{\tau_{fast}}e^{-t/\tau_{fast}} + \frac{B}{\tau_{slow}}e^{-t/\tau_{slow}} 
		\end{equation}
		
		The waveform profile of PET was fit with equation \ref{2exp}, though the fast component of PET was an order of magnitude greater in intensity than the slow component.
		
		PEN was found to be well described by a single exponential function of the form:
		
		\begin{equation}
			\frac{A}{\tau_{fast}}e^{-t/\tau_{fast}} 
		\end{equation}
		
		implying dominance of a single photoluminescence mechanism in PEN.
		
\begin{figure}[]
			\begin{centering}
				\subfloat[]{\includegraphics[width=0.44\textwidth]{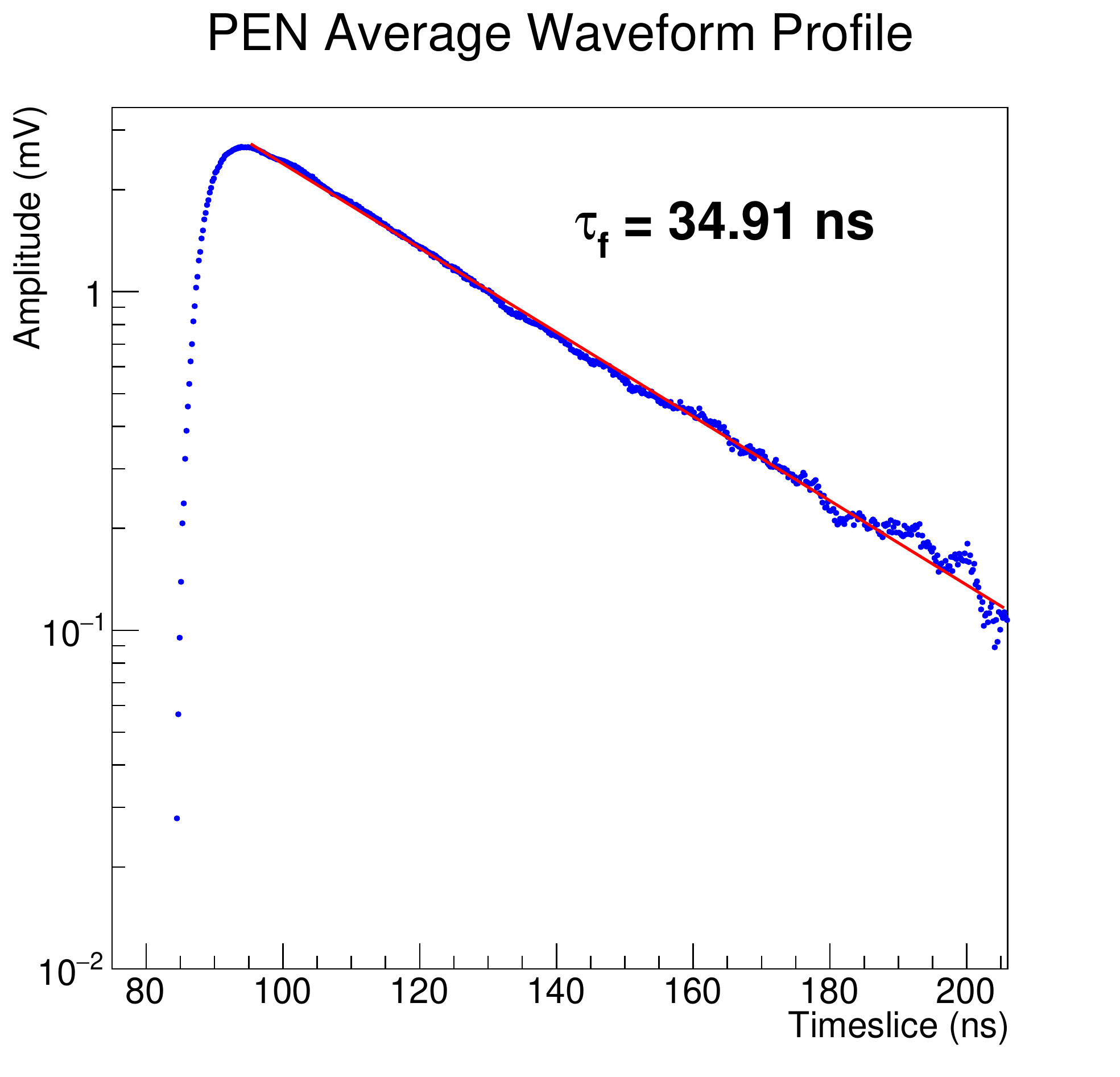}}
				\qquad
				\subfloat[]{\includegraphics[width=0.44\textwidth]{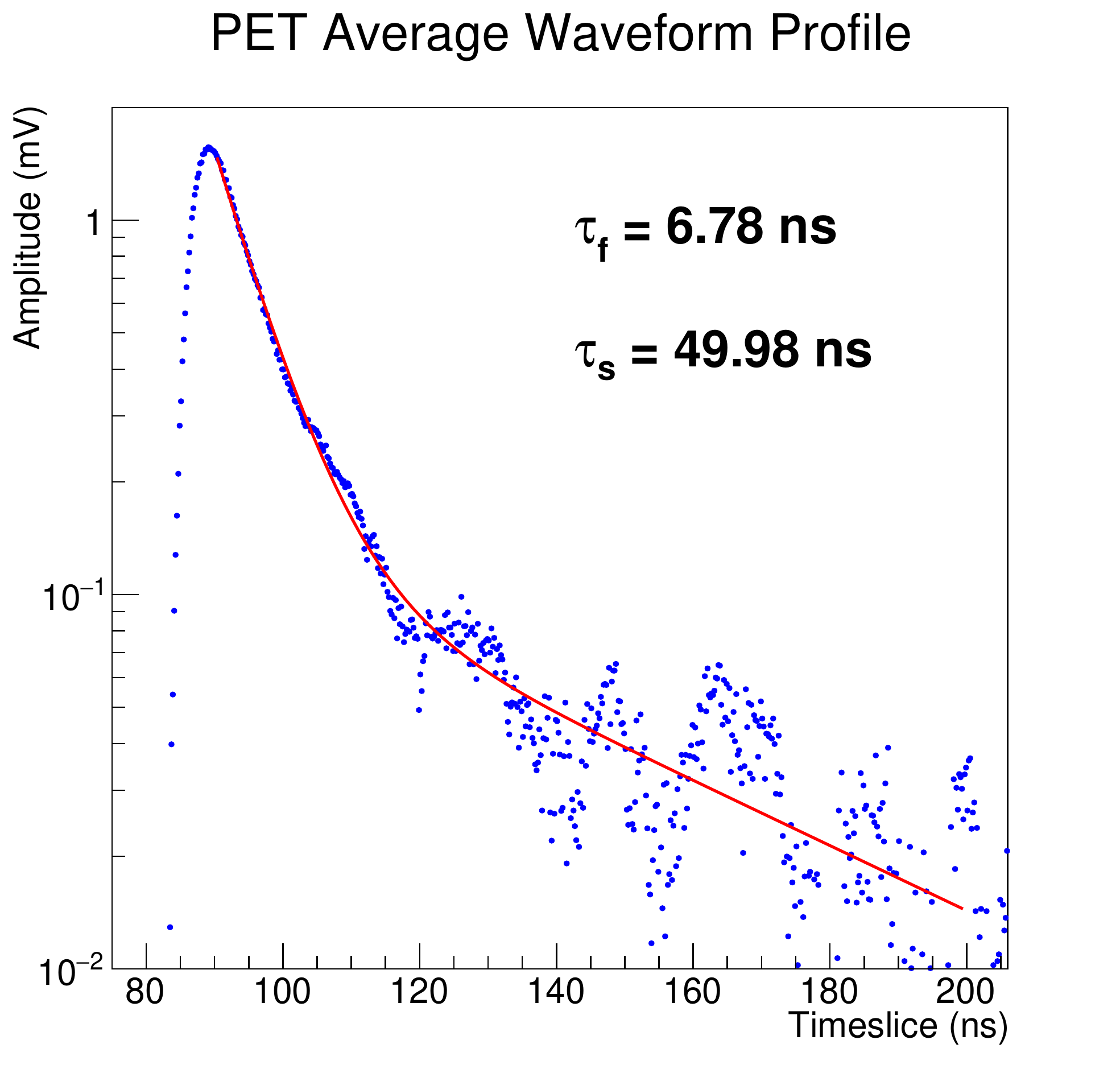}}
				\caption{Average waveform profiles with decay constants of PEN (a) and PET (b) using 120 GeV protons.}
				\label{figure:PENPETWF}
			\end{centering}
		\end{figure}

		Figure \ref{figure:PENPETWF} shows the waveform profiles and fits of PEN (a) and PET (b).  PEN and PET were found  to have dominant decay constants of 34.91 ns and 6.78 ns, respectively.

	\section{Conclusions}
		Waveform profiles of scintillating plastic sheets polyethylene naphthalate (PEN) and polyethylene teraphthalate (PET) were recorded by placing the scintillators in and perpendicular to an accelerated beam of 120 GeV protons.  The waveform profiles were measured using a PMT read out with an oscilloscope.  The exponential fits to the average waveforms indicate that PEN and PET have dominant decay constants of 34.91 ns and 6.78 ns, respectively.

	\section{Acknowledgements}
		The authors would like to thank the staff at Fermi National Accelerator Laboratory, in particular Mandy Rominsky, manager of the Fermilab Test Beam Facility.
	

\begin{thebibliography}{99}
	
	
	
	\bibitem{nakamura2011evidence}
	Nakamura, H., Shirakawa, Y., Takahashi, S., \& Shimizu, H. (2011). Evidence of deep-blue photon emission at high efficiency by common plastic. EPL (Europhysics Letters), 95(2), 22001.
	
	\bibitem{nakamura2013blended}
	Nakamura, H., Shirakawa, Y., Kitamura, H., Yamada, T., Shidara, Z., Yokozuka, T., ... \& Takahashi, S. (2013). Blended polyethylene terephthalate and polyethylene naphthalate polymers for scintillation base substrates. Radiation Measurements, 59, 172-175.
	
	\bibitem{JWandET}
	Wetzel, J., Tiras, E., Bilki, B., Onel, Y., \& Winn, D. (2016). Radiation damage and recovery properties of common plastics PEN (Polyethylene Naphthalate) and PET (Polyethylene Terephthalate) using a 137Cs gamma ray source up to 1.4 Mrad and 14 Mrad. Journal of Instrumentation, 11(08), P08023.
	
	\bibitem{evans2008lhc}
	Evans, L., \& Bryant, P. (2008). LHC machine. Journal of instrumentation, 3(08), S08001.
	
	\bibitem{DP}
	Scarpelli, A. (2018) ProtoDUNE and a Dual-Phase LArTPC, Neutrino Oscillation Workshop (NOW2018), 9 - 16 September, 2018, Rosa Marina (Ostuni, Brindisi, Italy), https://arxiv.org/pdf/1902.04780.pdf
	
	\bibitem{dualPhase}
	Cuesta, C. (2019) Status of ProtoDUNE Dual Phase, European Physical Society Conference on High Energy Physics - EPS-HEP2019 -10-17 July, 2019, Ghent, Belgium, https://arxiv.org/pdf/1910.10115.pdf
		
	\bibitem{antcheva2011root}
	Antcheva, I., Ballintijn, M., Bellenot, B., Biskup, M., Brun, et al. (2009). ROOT-A C++ framework for petabyte data storage, statistical analysis and visualization. Computer Physics Communications, 180(12), 2499-2512.
	
	\bibitem{swiderski2014measuring}
	Swiderski, L., Moszynski, M., Syntfeld-Kazuch, A., Szawlowski, M., \& Szczesniak, T. (2014). Measuring the scintillation decay time for different energy depositions in NaI: Tl, LSO: Ce and CeBr3 scintillators. Nuclear Instruments and Methods in Physics Research Section A: Accelerators, Spectrometers, Detectors and Associated Equipment, 749, 68-73.
	  
	
	\end{thebibliography}

\end{document}